\documentclass[sigconf]{acmart}

\AtBeginDocument{%
  \providecommand\BibTeX{{%
    \normalfont B\kern-0.5em{\scshape i\kern-0.25em b}\kern-0.8em\TeX}}}

\usepackage{enumitem}
\usepackage{bbding}
\usepackage{multirow}
\usepackage{colortbl}
\usepackage{setspace}
\usepackage{url}

\copyrightyear{2023} 
\acmYear{2023} 
\setcopyright{acmlicensed}\acmConference[DIS '23]{Designing Interactive Systems Conference}{July 10--14, 2023}{Pittsburgh, PA, USA}
\acmBooktitle{Designing Interactive Systems Conference (DIS '23), July 10--14, 2023, Pittsburgh, PA, USA}
\acmPrice{15.00}
\acmDOI{10.1145/3563657.3595989}
\acmISBN{978-1-4503-9893-0/23/07}

\acmSubmissionID{dis23-31}

\begin{document}

\title{Designing Conversational Multimodal 3D Printed Models with People who are Blind}

\author{Samuel Reinders}
\email{samuel.reinders@monash.edu}
\orcid{0000-0001-5627-413X}
\affiliation{
  \institution{Monash University}
  \city{Melbourne}
  \state{Victoria}
  \country{Australia}
}

\author{Swamy Ananthanarayan}
\email{swamy.ananthanarayan@monash.edu}
\orcid{0000-0002-9808-5844}
\affiliation{
  \institution{Monash University}
  \city{Melbourne}
  \state{Victoria}
  \country{Australia}
}

\author{Matthew Butler}
\email{matthew.butler@monash.edu}
\orcid{0000-0002-7950-5495}
\affiliation{
  \institution{Monash University}
  \city{Melbourne}
  \state{Victoria}
  \country{Australia}
}

\author{Kim Marriott}
\email{kim.marriott@monash.edu}
\orcid{0000-0002-9813-0377}
\affiliation{
  \institution{Monash University}
  \city{Melbourne}
  \state{Victoria}
  \country{Australia}
}

\renewcommand{\shortauthors}{Reinders, et al.}

\begin{abstract}
3D printed models have been used to improve access to graphical information by people who are blind, offering benefits over conventional accessible graphics. Here we investigate an interactive 3D printed model (I3M) that combines a conversational interface with haptic vibration and touch to provide more natural and accessible experiences. Specifically, we co-designed a multimodal model of the Solar System with nine blind people and evaluated the prototype with another seven blind participants. We discuss our journey from a design perspective, focusing on touch, conversational and multimodal interactions. Based on our experience, we suggest design recommendations that consider blind users' desire for independence and control, customisation, comfort and use of prior experience.
\end{abstract}



\begin{teaserfigure}
  \includegraphics[width=\textwidth]{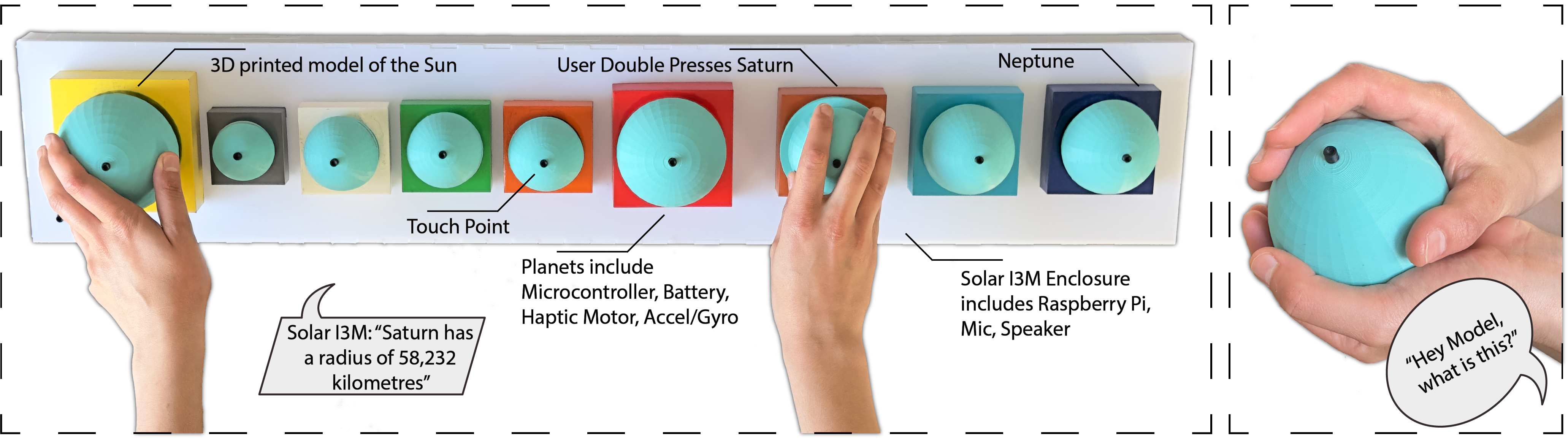}
  \caption{Solar I3M is designed to be touched, held, or picked up. The eight planets and the Sun sit upon an acrylic enclosure. Blind users can perform touch gestures on the planets to extract audio labels, or they can engage in conversational dialogue. Solar I3M also uses haptic vibratory feedback to add additional context during interactions.}
  \Description{Two photos of Solar I3M. The first, to the left, shows the entire Solar I3M model: the acrylic enclosure and 3D printed planets and Sun that sit on it. A person's hands are on top of Solar I3M, touching the Sun and Saturn printed components. This photo is also annotated, drawing attention to a touch point found on top of Mars, and highlighting that the planet models each include a microcontroller, battery, haptic motor and accelerometer. There is a speech bubble coming from Solar I3M, representing speech output of ``Saturn has a radius of 58,232 kilometres''. The second photo, to the right, shows a user holding a planet model and asking Solar I3M the question, ``Hey Model, what is this?''}
  \label{fig:teaser}
\end{teaserfigure}


\maketitle

\section{Introduction}
People who are blind experience difficulty accessing graphical information, particularly educational content~\cite{butler2017understanding,Phutane2021}. 3D printed models are becoming increasingly used as a technology for providing accessible graphics, with widespread interest amongst blind users in how they can be applied in everyday life. These models have benefits over traditional tactile graphics, including better representation of three-dimensional concepts, most notably in science education.

However, much like tactile graphics, 3D printed models often include braille labelling and legends, which can be problematic. Braille labels can occupy considerable space, limiting the amount of information that can be conveyed directly on the model. Separate legends can be provided, but these add to cognitive load as blind users may have to constantly shift between the model and the legend~\cite{Holloway2018}. Moreover, the use of braille is not ideal due to limited braille literacy among blind users~\cite{NFIB2009}. As a result, researchers have developed interactive 3D printed models (I3Ms) with pre-recorded audio labels~\cite{Shi2016,Reichinger2016,Giraud2017,Gotzelmann2017,Holloway2018,Ghodke2019,Davis2020}. Although these I3Ms are an improvement, they offer limited passive interactivity, consisting only of basic contextual information retrieval. 

Recent research has identified that blind users want rich interactions with I3Ms in ways similar to their personal technology, using modalities such as touch and conversational interfaces~\cite{Reinders2020}. Our current work presents the next step in this nascent area, with the co-design of an I3M of the Solar System -- Solar I3M. Building on earlier works that elicited I3M interaction modalities and strategies directly from blind end-users~\cite{Reinders2020, Shi2017b}, our work explores the creation of a fully functional multimodal 3D model that includes touch interaction, haptic vibratory feedback and a conversational interface (Figure \ref{fig:teaser}). Solar I3M was developed with people who are blind across multiple co-design sessions, and shares a tight integration of modalities. Our contributions include:

\begin{itemize}
    \item A co-design focused account of Solar I3M and its support for richer multimodal interactions that enable more natural interaction paradigms. To the best of our knowledge, this is the first I3M for blind users that includes a conversational interface (with demonstrative pronouns) associated with a physical tangible model.
    \item A usability evaluation of Solar I3M that revealed that blind users share a desire for richer multimodal interactions that are physically embodied and the ability to customise these according to their personal preferences.
    \item Design recommendations based on our co-design process and evaluation to help facilitate future I3M research and production. Particularly important was identifying how key areas of functionality -- touch, conversational and multimodal interaction -- could be designed to better support blind users' desire for: use of \textit{prior experience}, interactions that uphold their \textit{independence and control} and respect their sense of \textit{confidence and comfort}, and allow \textit{customisation and personalisation}.
\end{itemize}

The enthusiastic evaluation of Solar I3M by blind users demonstrates the potential of these kinds of models in an educational context. They could also be used in museums and other cultural institutions. We therefore believe that our research and Solar I3M will be of interest to practitioners creating accessible education products and accessible exhibits at museums and other cultural institutions. 

\section{Related Work}
\subsection{Accessible Graphics \& 3D Printed Models}
Accessing graphical information can prove difficult for people who are blind, impacting education opportunities~\cite{butler2017understanding} and making independent travel difficult~\cite{Sheffield2016}. These can lead to reductions in confidence and quality of life~\cite{Keeffe2005}. 
Tactile graphics are used by blind people to assist in classroom learning~\cite{Aldrich2001,Rosenblum2015} and orientation and mobility training~\cite{Blades1999,Rowell2005} and are commonly provided with braille labels and legends. However, they can only convey limited amounts of contextual information, chiefly due to the amount of space that braille requires~\cite{brock2012,Holloway2018,Shi2016}. Additionally, conventional production methods can only convey height/depth to a limited degree~\cite{Holloway2018}, restricting the types of graphics that can be produced. 

3D printed models have increasingly become an alternative, allowing a broader range of graphics to be produced when compared to conventional tactile graphics, including more complex three-dimensional concepts. The cost and effort involved in producing 3D printed models have fallen more in line with that of tactile graphics, resulting in 3D printing being studied broadly across many accessible graphics application areas: mapping and navigation~\cite{gual2012visual,Holloway2018,Holloway2019b,Holloway2022}; special education~\cite{Buehler2016}; books~\cite{kim2015,Stangl2015}; mathematics~\cite{Brown2012,Hu2015}; graphic design~\cite{McDonald2014}; science~\cite{wedler2012applied,Hasper2015} and programming curricula~\cite{kane2014}. While 3D printed models offer benefits over tactile graphics, braille labelling remains an issue: the low-fidelity of 3D printed braille can limit the readability of braille labels~\cite{Brown2012,Taylor2015,Shi2016}, there is limited space for labels and updates and additions to braille require model reprinting~\cite{Holloway2018}. Additionally, as with tactile graphics, reliance on braille labels is not ideal for the majority of blind people who have limited braille literacy~\cite{NFIB2009}.

\subsection{Interactive 3D Printed Models (I3Ms)}
To help overcome labelling limitations, 3D printed models are increasingly being combined with low-cost electronics and smart devices to produce interactive 3D printed models (I3Ms). I3Ms have been created and applied across many blind-specific contexts: mapping and navigation~\cite{Gotzelmann2017,Holloway2018,Shi2020}; art~\cite{Holloway2019,Bartolome2019}; and education~\cite{Ghodke2019,Shi2019,Reinders2020}. Many I3Ms include button or touch-triggered audio labels that when activated describe different details of the model~\cite{Shi2016,Reichinger2016,Giraud2017,Gotzelmann2017,Holloway2018,Ghodke2019,Davis2020}. Audio labels can be stored as either pre-recorded files or synthesised from text in real-time. 
Audio labels are particularly useful for blind users who are not braille readers, and the label information can be updated easily.

The design space of I3Ms is still very much in its infancy, with much of the current research in the field focused on simple touch interactions and extraction of basic information using passive audio labelling. Recent works using more user-driven design approaches have begun to uncover a desire amongst blind users for richer, more interactive experiences~\cite{Shi2017b,Shi2019,Reinders2020}. One study used a Wizard-of-Oz (WoZ) method and identified that blind users want to interact with I3Ms across three modalities -- gestures, speech, and buttons~\cite{Shi2017b}. Further work found that, in addition to passive auditory labelling, I3Ms should allow blind users to ask questions using more conversational language~\cite{Shi2019}, while our earlier work found that blind users want to interact in ways similar to their personal technology~\cite{Reinders2020}, including the use and combination of touch gestures, conversational dialogue, and haptic vibratory feedback in more multimodal interactions.

\subsection{Conversational Interfaces}
People who are blind find voice interaction convenient~\cite{Azenkot2013}, and adoption rates of devices offering conversational interfaces -- e.g. Siri (Apple), Alexa (Amazon) and Google Assistant -- are high amongst this group~\cite{Pradhan2018}. Recent work into the accessibility of conversational interfaces integrated into devices like smartphones and smart speakers has found that they often assume a strict human-to-human conversational model that can limit the interactions of blind users~\cite{Abdolrahmani2018,Branham2019}. ~\cite{Choi2020} found that the slow speed of conversational output might conflict with the speech comprehension rates of many blind users, with other research identifying a lack of customisation options relating to the voice used by conversational interfaces, and response length~\cite{Abdolrahmani2018,Branham2019}.

Little work has been undertaken exploring how conversational interfaces can be integrated into I3Ms. Such models could allow blind users to engage and ask questions using natural language, allowing for more natural interaction paradigms and richer interactive experiences. ~\cite{Quero2019} combined a tactile graphic with a conversational interface focused on indoor navigation, but voice interactions were performed through a connected smartphone rather than the model itself. ~\cite{Bartolome2019,Quero2018} created voice-controlled guides to help blind users navigate art pieces, however, their implementations supported simple and rudimentary commands rather than more expansive conversational dialogue. Other work has identified, but not comprehensively designed or tested, a set of I3M conversational interactions. ~\cite{Shi2019} suggested allowing blind users to extract information using conversational language, while our previous work showcased a simple WoZ implementation of wake words that could be used to invoke an I3M conversational interface~\cite{Reinders2020}.

\subsection{Multimodal Interfaces}
The provision of multiple modalities can increase the adaptability of a user interface~\cite{Reeves2004}, the resolution of information that the interface can communicate, and allow more natural interaction~\cite{Bolt1980}. Adaptability is particularly important for blind users, affording choice on how they interact. For example, they might choose to interact with an interface that offers both voice and touch input differently based on their ability, environment or even level of comfort~\cite{Abdolrahmani2018}. Richer resolution of information is also important for blind users, as the combination of accessible modalities -- e.g. tactile and auditory output -- may more closely match the capacity of vision and help overcome the \textit{``bandwidth problem''}~\cite{Edwards2015}. Multiple modalities can also improve the confidence and independence of blind users~\cite{Quero2021}. 
Modalities can also be combined to create more \textit{``natural user modalities''}, as described by the \textit{``Put-That-There''} paradigm ~\cite{Bolt1980}. This work combined gestural and speech inputs using demonstrative pronouns, and is foundational to the gestural-based interactions seen in virtual and augmented reality~\cite{Mayer2020}. 

Multimodal accessible aids have been created for blind users combining tactile graphics or touch screens with auditory output~\cite{Poppinga2011,Hamid2013,Quero2021}, haptic vibrations~\cite{Petrie2002,Goncu2011,Poppinga2011}, olfactory and gustatory perception~\cite{Brule2016}, and visual feedback for low-vision users that still possess levels of residual vision~\cite{Albouys2018}. I3Ms are multimodal and largely combine physical models with auditory output, albeit in a rather limited fashion~\cite{Gotzelmann2017,Holloway2018,Quero2019,Ghodke2019,Thevin2019,Davis2020}. However, our previous work found that blind users desired to also engage with haptic vibratory output, desiring seamless combinations of touch, haptic vibratory feedback, and conversational dialogue~\cite{Reinders2020}.

\subsection{Motivation}
With much of the work into I3Ms being preliminary in nature, often dealing with hypothetical or limited prototypes, we lack guidelines on how I3Ms should be designed. In particular, the design of I3Ms that integrate multiple modalities and make use of conversational interfaces is virtually unexplored despite the potential of such interfaces to support richer and more natural interaction.

\section{Solar I3M: A Co-Design Journey}
\label{sec:codesign}
\subsection{Overview}
To address this gap we designed and evaluated a fully functional multimodal I3M that supports touch gesture interaction, haptic vibratory feedback, audio labels, and a conversational interface. We chose to build a model of the Solar System. This not only reflects the high demand by blind students for more accessible STEM materials~\cite{butler2017understanding} and the rich opportunities for multimodal interaction revealed in our previous work~\cite{Reinders2020}, 
but also affords us the opportunity to contrast our design findings with earlier work where we used a partial-WoZ I3M implementation.

Importantly, we co-designed Solar I3M with multiple blind co-designers. The use of more participant-oriented methods, like co-design, is imperative when designing for and working with people with disability, 
as the researchers may not possess lived experience of the relevant disability or impairment. Co-design has been used in the blind accessibility space~\cite{Ghodke2019,Giraud2020,Nagassa2023}, and such methods have been recommended to address the low involvement of blind users across the design and evaluation of accessible graphic research~\cite{Butler2021,Kim2021}. This is particularly important as the degree of how fit-for-purpose accessible aids/tech are can influence abandonment rates~\cite{Betsy1993}. 

We held two design workshops, each with different groups of blind co-designers, and two one-on-one design sessions with a single blind expert co-designer (Table \ref{tab:table3}). Recruited through our lab's contact pool, our S1 \& S2 co-designers had no prior co-design experience, while our expert co-designer, involved in S3 \& S4, had prior design experience from unrelated research projects. 

\begin{table}[h!]
\caption{Outline of the four co-design sessions}
\Description{Four co-design sessions were held. Session 1 consisted of four users and focused on touch gestures. Session 2 consisted of four different users and focused on audio labelling, the conversational interface and haptic vibration. Sessions 3 and 4 were held with one unique co-designer focusing on all functionality  areas of Solar I3M's design. All four design sessions lasted 90 minutes.}
\label{tab:table3}
\begin{tabular}{|c|c|c|p{40mm}|}
\hline
\rowcolor[HTML]{EFEFEF} 
\textbf{Session} & \textbf{Users} & \textbf{Duration} & \textbf{Session Focus} \\ \hline
S1 & 4 & 90mins & Touch Gestures \\ \hline                              
S2 & 4 & 90mins & Audio Labelling \newline Conversational Interface \newline Haptic Vibration  \\ \hline
S3 & 1 & 90mins & Touch Gestures \newline Audio Labelling \newline Conversational Interface  \\ \hline
S4 & 1 & 90mins & Audio Labelling \newline Conversational Interface \newline Haptic Vibration \\ \hline
\end{tabular}
\end{table}



\begin{figure*}[h!]
\centering
\includegraphics[width=40pc]{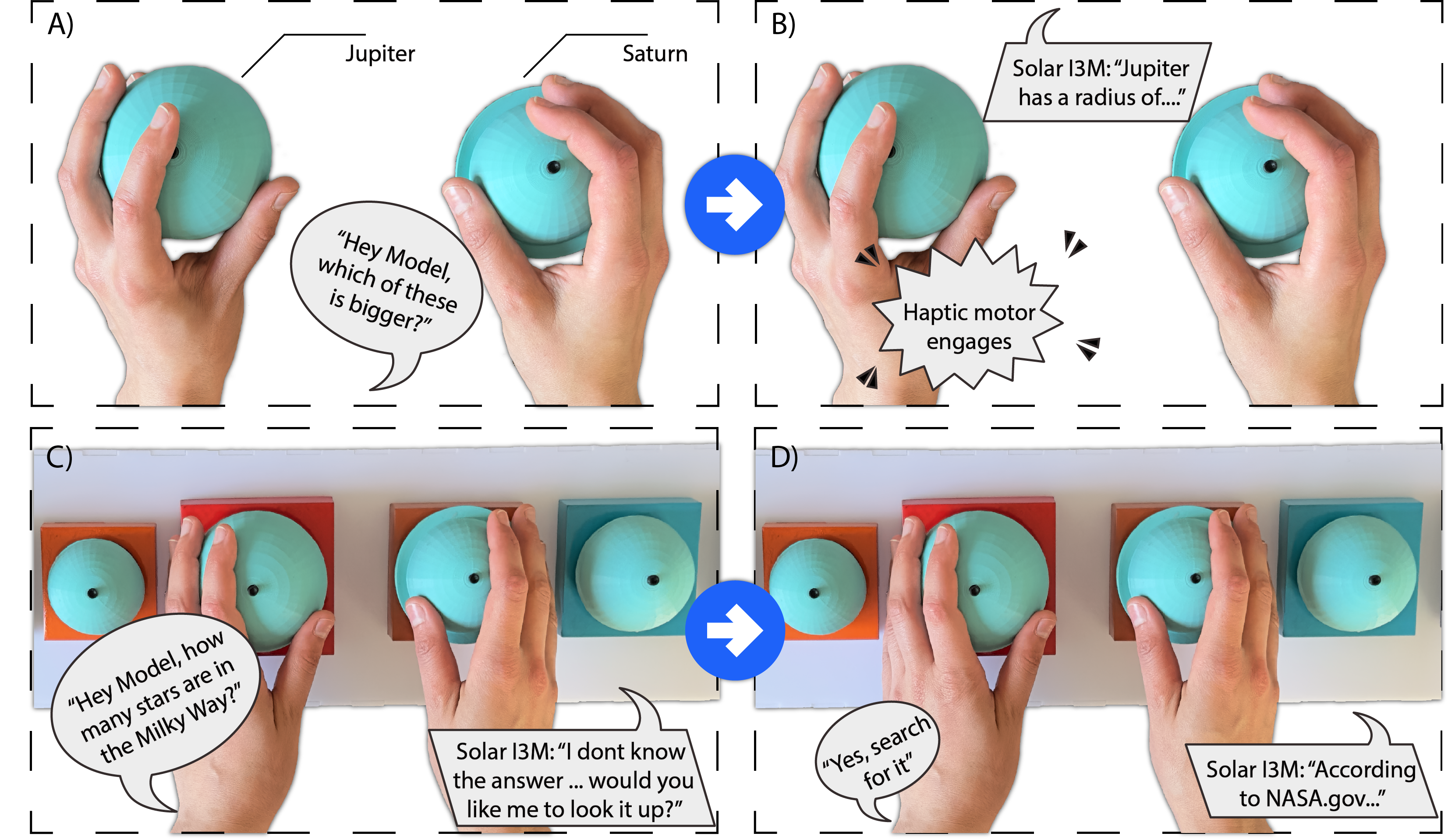}
\caption{Solar I3M allows blind users to engage in rich multimodal experiences. A) Users can engage in conversational dialogue using wake words and demonstrative pronouns; B) Solar I3M combines auditory responses simultaneously with haptic vibratory feedback; C \& D) Users can perform searches to extend their knowledge.}~\label{fig:sample-interaction}
  \Description{Four photos of Solar I3M interactions. The first interaction encompasses two images, A) \& B). A) shows a user holding two planets in their hands, Jupiter and Saturn, asking Solar I3M the question ``Hey Model, which of these is bigger?''. B) shows the next stage of this interaction, with Solar I3M responding ``Jupiter has a radius of...', with attention drawn to a haptic vibration that simultaneously triggers from the Jupiter model. The second interaction showcases how Solar I3M can search the internet across C) \& D). C) shows a user placing their hands on Jupiter and Saturn, asking Solar I3M how many stars are in the Solar I3M, receiving the response ``I don't know the answer, but I can look it up''. D) shows this interaction being completed, with the user authorising the search and subsequently receiving the correct answer from Solar I3M}
\end{figure*}

Solar I3M's design was inspired by prior work eliciting hypothetical I3M interaction modalities from blind users using WoZ experiments~\cite{Shi2017b,Reinders2020}. What makes Solar I3M unique, in addition to being the first I3M with a working conversational interface integrated directly into the model, is that conversational dialogue, audio, haptic vibration, and touch are tightly integrated to deliver a rich multimodal experience. 
The following illustrates a typical interaction (Figure \ref{fig:sample-interaction}):
\begin{enumerate}
\item A blind user picks up 3D printed models of Jupiter and Saturn, gathering details like shape and size tactually.
\item They proceed to press the touch points on each planet; Solar I3M identifies them via an audio label using an attached speaker --  \textit{``Activated Jupiter'', ``Activated Saturn''}.
\item Continuing, the user engages in conversation, asking Solar I3M a question -- \textit{``Hey Model, which of these is bigger?''}.
\item Solar I3M provides the spoken response -- \textit{``Jupiter has a radius of 69,911 kilometres, while Saturn has a radius of 58,232 kilometres''} -- emitting haptic vibrations from each planet model as they are referred to.
\item The user continues their interaction, asking \textit{``Hey Model, how many stars are in the Milky Way?''}.
\item Solar I3M replies -- \textit{``I don't know the answer to: `How many stars are in the Milky Way', but would you like me to search for it?''} -- to which the user responds -- \textit{``Yes, search for it''}.
\item Solar I3M speaks -- \textit{``According to NASA.gov, there are over 100 billion stars in the Milky Way''} -- and the user decides to conclude their current interaction.
\end{enumerate}

Our co-design journey is broken down around three areas -- \textbf{touch, conversational and multimodal interaction} -- highlighting how our blind co-designers and existing research informed Solar I3M's design. 

\subsection{Co-Designing Touch Interactions}
Like traditional tactile graphics, I3Ms are designed to be held and touched, allowing blind end-users to undertake information gathering through tactile exploration. What makes I3Ms unique is that in addition to touch being used to explore an object, touch can be used as an input to trigger I3M functionality. Many I3Ms make use of touch to trigger audio labels that playback audio labels~\cite{Shi2016,Giraud2017,Holloway2018,Ghodke2019,Davis2020}. However, unlike
smartphones, where the enclosure is gripped and gestures performed on a dedicated touch surface, I3Ms are held
and tactually explored uninterrupted, with touch points sharing the same faces of the object being actively explored.
This presents a design challenge, with research showing that blind participants can become frustrated during I3M tactile
exploration if interrupted by the accidental activation of system output~\cite{Holloway2018,Reinders2020}.

\subsubsection{Creating Easily Identifiable Touch Points}
\begin{figure}[h]
  \centering
  \includegraphics[width=1.00\columnwidth]{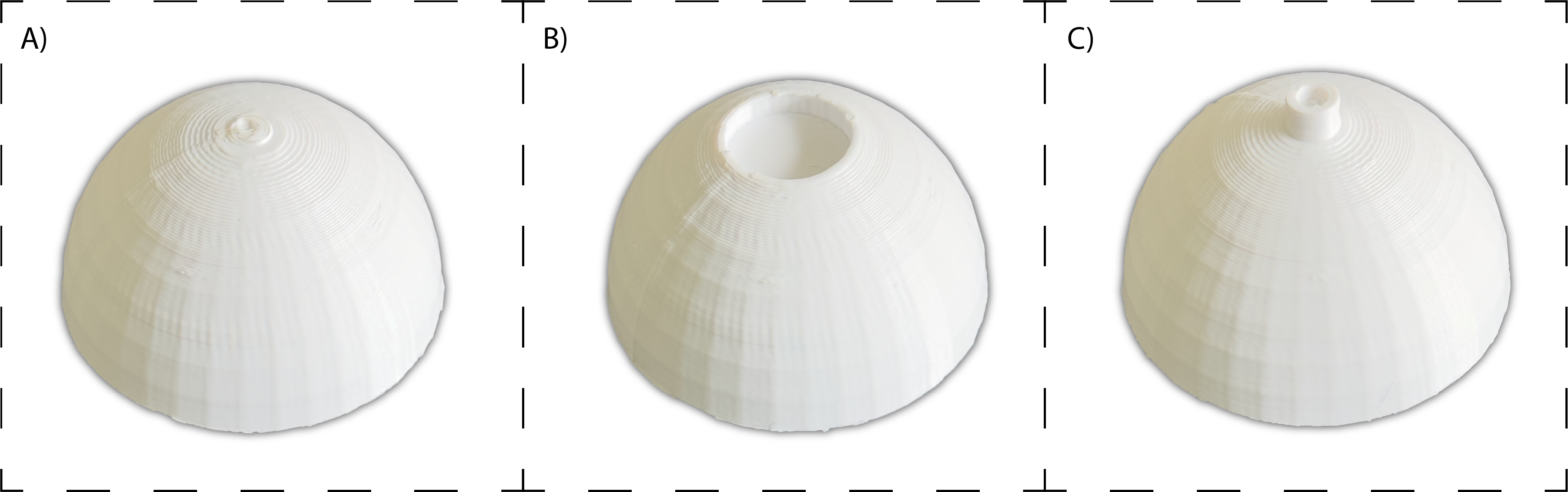}
  \caption{Touch point designs from S1 design session. A) tappable area integrated flush; B) indented touch surface; C) raised touch point.}~\label{fig:touch-points-one}  
  \Description{Three photos of the touch point designs that were given to co-designers during the S1 design session. The first photo shows a 3D printed semisphere where the touch surface is integrated flush into the print at the northern pole. The second design shows a semisphere where there is a small circular indent on the northern pole where the touch surface sits. The third design, which was chosen by our co-designers, is a semisphere with a raised touch point design that protrudes out of the northern pole like a small tower approx 5mm.}
\end{figure}

Using electrically conductive printing filament, it is possible to create 3D prints where specific features of areas of the print are conductive. When connected to capacitive touch sensors, these areas can act as tappable surfaces or touch points. During S1 we explored designing easily discoverable touch areas. Our co-designers were handed three 3D printed planet models (Figure \ref{fig:touch-points-one}): (1) a planet with a tappable area integrated flush into the surface of its northern pole, (2) another with a small indented touch surface at the top of the planet, and (3) a planet with a discrete raised touch point that protruded out of the northern pole. The first two designs were influenced by an existing recommendation that touch points shouldn't distort the surface of a print~\cite{Holloway2018}. We asked our co-designers which designs were most distinctive. Three co-designers gave immediate preference towards the protruded design, suggesting that the shape, size, and feel of the touch point made it easier to locate amongst other details of the printed model. 
The protruded design deviates from ~\cite{Holloway2018}'s guideline, 
suggestive that this recommendation might be specific to domains requiring a greater degree of touch exploration, e.g. I3Ms of maps that have more salient tactual details. 

\subsubsection{Choosing Touch Gestures \& Their Uses}
\label{sec:gestures}
Touch gestures are well understood due to the proliferation of touch input devices, e.g. smartphones, with high rates of use amongst blind users. ~\cite{Holloway2018} implemented Single, Double and Long touches in their I3M to playback audio labels, while ~\cite{Reinders2020} observed participants designing their own interactions based on smartphone analogous touch gestures. We drew from these, implementing touch gestures used in end-users' personal technology as follows:
\begin{itemize}
    \item \textbf{Single Tap/Press}: Identifies the object being interacted with using an audio label, e.g. \textit{``this is Jupiter''};
    \item \textbf{Double Tap/Press}: Rotates a selection of audio label facts about the object being interacted with -- colour, size, order, distance/location, composition, and number of moons, e.g. \textit{``Mars is reddish brown, with no clouds, and has white ice caps''};
    \item \textbf{Long Tap/Press}: Engages Solar I3M's conversational interface.
\end{itemize}

\begin{table}[h!]
\caption{Touch gestures supported by Solar I3M}
\Description{Three different touch gestures were designed. Activate/Single Press requires a touch point to be held for 500ms. The first press Activates a touch point, and subsequent presses identify that touch point. Double Press requires a touch point to be pressed twice within 500ms. Performing this gesture rotates through a series of touch point audio-labelled facts. The final gesture, Long Press, requires users to press a touch point for 2 seconds. The Long Press engages Solar I3M's conversational interface.}
\label{tab:table4}
\begin{tabular}{|c|c|p{35mm}|}
\hline
\rowcolor[HTML]{EFEFEF} 
\textbf{Gesture} & \textbf{Timing}  & \textbf{Output} \\ \hline
Activate/Single Press & 500ms & First press activates a touch point, subsequent press identifies touch point  \\ \hline
Double Press  & 500ms & Rotates through a series of touch point facts      \\ \hline
Long Press & 2s    & Engages Solar I3M's conversational interface       \\ \hline
\end{tabular}
\end{table}

\subsubsection{Ensuring That Touch Interactions Are Deliberate}
Initially, Solar I3M's touch gestures were influenced by existing standards~\footnote{https://developer.apple.com/design/human-interface-guidelines/inputs/touchscreen-gestures/}, for consistency with personal technology use. A Single Tap was designed to trigger after a touch event of 250ms had been performed on a touch point, a Double Tap when two touch events were detected within 250ms, and a Long Tap for touch events lasting longer than 500ms. 

During S1 we asked our four co-designers to repeatedly perform each touch gesture on the touch point of one of Solar I3M's printed planets. All co-designers experienced problems reliably performing the touch gestures, including Double Taps being detected as successive Single Taps. One co-designer became visibly irritated during this activity and indicated that the touch gestures made them feel frustrated, as they found the timing windows too brief/precise. With the detection timing windows of gestures adjustable on the spot, we took turns with our co-designers adjusting the timing of each touch gesture until a setting was found that our co-designers could reliably and consistently perform. The timing of each gesture was increased (Table \ref{tab:table4}), 
suggesting that blind users may require additional time and concentration when performing touch gestures on I3Ms compared to personal technology with dedicated touch surfaces.

We returned to touch gestures in S3, and while our co-designer was able to reliably perform all of the touch gestures, they suggested improvements:
\begin{itemize}
\item \textbf{Touch gesture names should reflect how they are performed}. Touch gestures required longer presses compared to the quicker taps they performed on their smartphone. Our co-designer suggested that referring to touch gestures as \textit{Presses} rather than \textit{Taps} would be more illustrative (Table \ref{tab:table4});
\item \textbf{Touch points should require activation}. While feeling in control during tactile exploration, our co-designer suggested that an initial \textit{Single Press} should be required to \textit{activate} a touch point before subsequent touch gestures could be performed. They detailed that this would increase their likelihood of engaging in tactile exploration as it would reduce inadvertent interruptions from touch gesture-driven audio label playback. A parallel was made to the screen reader built into the co-designers smartphone which requires a gesture to activate the currently selected interface element.
\end{itemize}

\subsubsection{Design Outcomes}
After S1 \& S3, we had a clear outline of how our blind co-designers wanted to undertake touch interactions. \textbf{Independence during tactile exploration} was important, allowing users to freely explore Solar I3M without the interruption of accidental audio label playback. Maintaining a \textbf{sense of control during touch interactions} was also seen as crucial, supporting users in clearly identifying which areas of  Solar I3M touch interactions can be performed, and ensuring that touch gestures work reliably and with confidence.

\subsection{Co-Designing Conversational Interactions}
Conversational interfaces allow blind users to navigate and complete tasks in a non-visual manner. Despite their widespread use by people who are blind, recent work investigating the accessibility of mainstream conversational interfaces has found that their design can limit blind users' interactions, as dialogue is based on human-to-human conversational models~\cite{Abdolrahmani2018,Branham2019}. 

Little work has been undertaken exploring how conversational interfaces can be integrated into I3Ms. ~\cite{Quero2019} paired a conversational interface with an I3M to assist blind users with indoor navigation. However, their conversational interface was activated through a connected smartphone rather than the I3M itself. ~\cite{Shi2019} suggested that blind students should be able to ask I3Ms questions about modelled content, while ~\cite{Reinders2020} found that blind users wanted to talk directly to I3Ms using natural language to fill gaps in their knowledge using wake words. 

We built Solar I3M's conversational interface using Dialogflow\footnote{https://cloud.google.com/dialogflow} and its Python API\footnote{https://github.com/googleapis/python-dialogflow}, and drew from prior work into the accessibility of mainstream conversational interfaces.

\subsubsection{Customising Conversational Output}
During S2 \& S3, we ran a series of activities to understand how our co-designers wanted to customise responses, and made adjustments on-the-fly using Speech Synthesis Markup Language (SSML).

During S2, we played back a series of audio labels generated at speeds comparable to standard human conversation. We asked our co-designers if there were any scenarios where they would like to change the speed of auditory output. Our co-designers suggested that using Solar I3M to revise content, or in a public space, that it should speak faster. One co-designer indicated that they would like Solar I3M to speak twice as fast. We modified the SSML of the audio label responses, generating versions at 150\% and 200\% speed. Our co-designers liked the faster-spoken responses, indicating that use might vary according to environment. The speed of Solar I3M's auditory responses can be changed using co-designer suggested voice commands -- \textit{``Can you speak faster [or] slower?''} and \textit{``Can you speak at [100/150/200\%] speed?''} (Table \ref{tab:table5}).

The synthesised voice that Solar I3M's conversational interface employs can also be changed. One of our co-designers indicated that the initial voice had an unpleasant quality, and sounded robotic. To explore this further, we opened the Google Text-to-Speech web page and began generating speech with different vocoders\footnote{https://cloud.google.com/text-to-speech/docs/voices}. All four of our co-designers found speech generated using the WaveNet vocoders, which use neural networks to synthesise  more human-like speech, more pleasant. One co-designer suggested that they would like to select different types of voices, similar to how Siri offers a selection of different voices. After the design session concluded, we implemented two WaveNet voices -- masculine and feminine -- to complement the existing Standard vocoder synthesised voice. The voice Solar I3M uses to generate auditory responses can be changed using the co-designer suggested voice command -- \textit{``Can you speak with a [feminine/masculine/robotic] voice?''} (Table \ref{tab:table5}).

The length and detail of Solar I3M's auditory responses can also be modified, influenced by ~\cite{Abdolrahmani2018}'s finding that blind users can find responses verbose or insufficiently detailed. During S3, we asked our one-on-one co-designer whether the length of a typical response -- e.g. \textit{``Mars has a radius of 3389 kilometres''} -- could be increased meaningfully. This co-designer felt that additional detail should only be added if it provided greater context and meaning, e.g. how the radius of Mars compared to Earth. We provided a more detailed response -- \textit{``Mars has a radius of 3389 kilometres, making it around half the size of Earth''} -- which our co-designer found useful, especially for objects or concepts they were learning about for the first time. After the session, we modified all auditory responses to have two variants -- \textit{low} and \textit{high} verbosity -- adjusted using the voice command -- \textit{``Can you give me [longer/shorter] answers?''} (Table \ref{tab:table5}).

\begin{table}[h!]
\caption{Users can modify Solar I3M's auditory responses based on voice, speed and length}
\Description{Users can customise how Solar I3M's auditory responses are generated. There are three voices - Feminine, Masculine and Robotic. These voices have three adjustable speeds - 100, 150, 200 - and all responses have short and long variations.}
\label{tab:table5}
\begin{tabular}{|c|c|c|c|}
\hline
\rowcolor[HTML]{EFEFEF} 
\textbf{Voice} & \textbf{Vocoder} & \textbf{Speed} & \textbf{Verbosity/Length} \\ \hline
Feminine & WaveNet & 100, 150, 200 & Shorter, Longer \\ \hline
Masculine & WaveNet & 100, 150, 200 & Shorter, Longer \\ \hline
Robotic & Standard & 100, 150, 200 & Shorter, Longer \\ \hline
\end{tabular}
\end{table}

\subsubsection{Ensuring That Users Are In Control During Interactions} 
When designing Solar I3M's conversational interface, we wanted to ensure that blind users could undertake conversational interactions that upheld and respected their sense of control. During S2, we ran a series of activities to understand how blind users wanted to invoke and engage with Solar I3M's conversational interface. 

We began by asking our co-designers what wake words they wanted to invoke the conversational interface with. We provided an example: \textit{``Hey Model''}. One co-designer thought that \textit{``Hey Model''} was memorable. Two of the other co-designers initially suggested that Solar I3M could use a more descriptive wake word, e.g. \textit{``Solar''} or \textit{``Solar System''}, but later indicated that using a more generalisable phrase, like \textit{``Hey Model''}, meant they wouldn't have to remember different wake words between I3Ms. We used Picovoice Porcupine\footnote{https://picovoice.ai/platform/porcupine/} to generate the wake word \textit{``Hey Model''}. Our co-designers were comfortable taking turns saying this phrase. 

We asked our co-designers if it was important to offer a manual method of activating Solar I3M's conversational interface. One co-designer drew connections to their smartphone, highlighting how they primarily activated their conversational interface by holding down a button, as they found it more reliable. This aligns with ~\cite{Abdolrahmani2018}'s finding
that manual forms of activation are important for blind users, as they can become quickly frustrated when wake words aren't properly detected. We outlined our \textit{Long Press} gesture (Section \ref{sec:gestures}), with one co-designer directly relating this to how they would hold down the \textit{Side Button} on their iPhone.

To further investigate the design of Solar I3M's conversational interface, during S4 we gave our co-designer ample time to initiate conversational interactions. While this co-designer enjoyed their time with  Solar I3M, they offered the following suggestions:

\begin{itemize}
\item \textbf{A voice command to repeat auditory outputs}. Our co-designer found that longer auditory responses, particularly those returned externally from Google Knowledge Graph (Section \ref{sec:searching}), were difficult to fully comprehend. They proposed a voice command that would repeat Solar I3M's most recent auditory response, suggesting the phrase -- \textit{``Can you repeat that?''};
\item \textbf{Longer recording windows}. Our co-designer suggested that the microphone timeout window of Solar I3M was too short and cut them off whilst they were still asking their question. We modified the default microphone timeout of the Dialogflow Python API, adjusting it so that Solar I3M would continually record a user's request until no further microphone input was detected; 
\item \textbf{Appropriate feedback when recording}. Initially, when Solar I3M's conversational interface was invoked, the microphone would automatically begin recording a user's speech. Our co-designer felt that additional feedback was required for two reasons: (1) to take the guesswork out of when they could begin their request, and (2) it made them feel more secure knowing when the I3M was listening to them. The phrases -- \textit{``Recording''} and \textit{``I am listening''} -- were suggested to help improve their sense of control over conversational interactions;
\end{itemize}

\subsubsection{Comprehensive Conversational Interactions}
\label{sec:searching}
Solar I3M's conversational interface uses Dialogflow to provide natural language understanding. Operating in the cloud, this instance has been trained on a corpus of information relating to the Solar System. This covers the same information and knowledge accessible through touch gesture-driven audio labels (Section \ref{sec:multimodal}). Dialogflow also allows users to ask questions, 
allowing richer interactions where Solar I3M can fulfil a wider variety of requests, and help users in filling gaps in their knowledge. If Solar I3M's conversational interface cannot directly answer a question, it will offer to perform a search by asking -- \textit{``I don't know the answer, but would you like me to search for it?''}. Users can authorise by responding -- \textit{``Yes, search for it''} -- activating Google Knowledge Graph \footnote{https://developers.google.com/knowledge-graph} to fetch a response from sources including Wikipedia, Wikidata and CIA World Factbook. 

During the S4 one-on-one session, we wanted to understand whether this feature would be meaningful, and how its design could support blind users. We prompted our co-designer to ask Solar I3M a series of questions that it had not been trained to answer. After Solar I3M successfully returned responses from Google Knowledge Graph, our co-designer indicated that being able to perform searches would help to increase their confidence that their interactions would be fulfilled, and offered the following suggestions: 

\begin{itemize}
\item \textbf{Repeat the user's question before searching}. Initially, Solar I3M would prompt a user to perform a search by asking -- \textit{``I don’t know the answer, would you like me to search for it?''}. Our co-designer felt it was important for this prompt to repeat their question, allowing them to confirm that the conversational interface had correctly interpreted and understood their request. The format -- \textit{``I don't know the answer to \textbf{[question]}...''} -- was suggested; 
\item \textbf{Appropriate references should be provided}. Our co-designer felt that presenting a source for all external responses would be integral, particularly if used in education contexts, allowing them to make their own judgement call on the trustworthiness of responses. The format -- \textit{``\textbf{According to [source]}, [searched result]...''} -- was discussed;
\item \textbf{Asking for more information}. After asking a question and receiving an answer, our co-designer suggested that users should be able to ask for additional information as required. The voice command -- \textit{``Can you tell me more?''} -- was suggested as a way of rotating through a planet's facts, and the conversational interface prompt -- \textit{``According to [source], [answer]. \textbf{Would you like to know more?}''} -- was discussed.
\end{itemize}

\subsubsection{Design Outcomes}
Our S2, S3 \& S4 co-designers clearly expressed the ways in which they wanted to engage in conversational interactions. This centred on \textbf{customisable conversational outputs}, allowing users to modify and control how Solar I3M's auditory responses are generated based on preference, ability and environment. Additionally, ensuring \textbf{a sense of control during conversational interactions} was important, allowing users to choose how they initiate dialogue -- using wake words or touch gestures -- and ensuring appropriate feedback is provided so that interactions can be effectively understood and completed.

\subsection{Co-Designing Multimodal Interactions}
As previously discussed, multimodal interaction offers several advantages to blind users: richer resolution of information~\cite{Edwards2015}, more natural interaction~\cite{Bolt1980}, and increased adaptability ~\cite{Reeves2004}. 

\subsubsection{Allowing Users To Choose How They Interact}
\label{sec:multimodal}
Between co-design sessions, we were particularly interested in making Solar I3M more adaptable, providing additional choices to users for them to determine how they could undertake interactions. To increase adaptability, we took the corpus that Solar I3M's conversational interface was trained on and made sure that all information accessible through the conversational interface could also be extracted through touch gesture-driven audio labels. Users are afforded choice in how they interact -- e.g. a user may not feel comfortable engaging in conversational dialogue in public spaces~\cite{Abdolrahmani2018}, or may simply be tired, and may prefer to use a touch gesture to access information played back using an audio label (excluding search results discussed in Section \ref{sec:searching}). 

\subsubsection{Facilitating More Natural Multimodal Interactions}
As early as the 1980s, the \textit{Put-That-There} paradigm~\cite{Bolt1980} described how speech and gestural inputs can be combined to converge on more \textit{``natural user modalities''}. Their system allowed the use of demonstrative pronouns -- e.g. \textit{``that''} -- to refer to the objects in speech commands when pointing fingers at shapes on a screen. Demonstrative pronouns are foundational to gestural-based systems in virtual and augmented reality~\cite{Mayer2020}. We were interested in how demonstrative pronouns could be used in conversational dialogue during touch exploration, or when one or more planet models were picked up and held -- e.g. \textit{``what is this?''} -- or compared. Demonstrative pronouns could enable blind end-users to engage more naturally in interactions that combine different modalities.
Mainstream conversational interfaces, such as Siri and Alexa, do not support demonstrative pronouns, but ~\cite{Lee2021} modified the Google Assistant SDK to support \textit{``this''} and \textit{``these''} pronouns in a WoZ application. 

During S2, we explained the concept of demonstrative pronouns, and had our co-designers complete a hypothetical interaction: holding and comparing two planets, and asking -- \textit{``which of \textbf{these} is bigger?''}. While not yet supported by Solar I3M, we asked our co-designers what they thought of the interaction, and whether it was natural. Two co-designers suggested that it would allow them to perform interactions with objects they had yet to identify. One continued mentioning that they would feel less stressed about having to remember the name of each object. Our co-designers suggested additional pronouns including -- \textit{``this, it} and \textit{that''}. 

After the session concluded, we embedded accelerometers/gyroscopes into each planet model to allow tracking of which planets were being physically manipulated or held. Solar I3M's conversational interface was modified to support the use of demonstrative pronouns in voice commands and conversational dialogue. To explore this further, during S3 we asked our co-designer to complete a series of comparison tasks using the demonstrative pronouns. They felt that referring to planets using pronouns made them more willing to engage in conversational interactions, as it made them more natural and human-like. They also suggested that demonstrative pronouns should have some persistence between interactions -- e.g. the ability to pick up Mars and ask \textit{``how large is \textbf{this}?''}, before placing Mars back in its stand, picking up another planet and asking -- \textit{``is \textbf{it} larger than \textbf{this} planet?''}. 

\subsubsection{Augmenting Interactions with Haptic Vibratory Output}
The majority of previous work into multimodal accessible aids has focused on combinations of tactile graphics/printed models with auditory output~\cite{Gotzelmann2017,Holloway2018,Thevin2019,Ghodke2019}. ~\cite{Reinders2020} found a desire for combinations with haptic vibratory feedback, however, this was untested. Haptic vibratory feedback has been used in touch display systems~\cite{Goncu2011,Poppinga2011}, but by their very nature, they are less tangible than physical I3Ms. During S2, S3 \& S4, we explored how our co-designers wanted to engage in multimodal interactions, and how Solar I3M could utilise combinations of auditory and haptic inputs/outputs. 
As the components of Solar I3M were designed to be picked up and held, we were particularly interested in combinations of haptic vibratory feedback and auditory output.

During S2, we handed our co-designers a printed planet model with an embedded haptic motor, demonstrating basic haptic vibrations by turning the motor on and off. We asked our co-designers if they could think of any situations where they would like haptic vibrations to be used. One co-designer expressed excitement and suggested that vibrations could be used to help identify specific objects of interest -- e.g. a particular planet. We designed for this during the session, so that whenever an auditory response made reference to a particular planet, a one-second haptic vibration would be emitted from that planet model. We asked our co-designers to find Jupiter, and upon receiving the auditory response -- \textit{``Jupiter is the fifth planet from the Sun''} -- with a haptic vibration emitting from Jupiter, all co-designers agreed that: 

\begin{itemize}
\item When engaging in tactile exploration, haptic vibrations combined with auditory responses would provide the information necessary to find objects or to gain an approximate location of where they could be found; 
\item The use of the haptic vibration helped to better convey the location of the planets in a non-visual form.
\end{itemize}

Progressing, one co-designer asked if they could perform their own interaction. They picked up Neptune and Venus, performed a \textit{Long Press} on Neptune, and asked -- \textit{``Do Neptune and Venus have any moons?''}. Receiving the auditory response -- \textit{``Neptune has 14 moons, including..., while Venus has 0 moons.''} -- this co-designer expressed confusion regarding the fact that haptic vibrations had triggered from both planet models at the same time, and suggested that haptic vibrations should be carefully timed and synced. We made this change during the session, and when our co-designer performed this interaction again, they agreed that the haptic vibrations added more meaning to their interactions, and helped to better relate information regarding any tangible objects being held.

During S4, we revisited haptic vibratory feedback. Our co-designer suggested that haptic vibrations could provide feedback to confirm any touch gestures performed. They indicated that it would be useful to know when performing a touch gesture if they had executed it correctly, connecting this to the use of their smartphone screen reader that emits small haptic vibrations in response to gestural inputs. During the session, we created three unique haptic vibrations of varying duration, each mapped to a specific touch gesture (Table \ref{tab:table6}). Resuming, our co-designer then performed various touch gestures on Mars, and indicated that the haptic vibrations made their interactions more engaging and that confirming their inputs helped give them a greater sense of control. 

\begin{table}[h!]
\caption{Each touch gesture triggers a unique haptic vibration}
\Description{Three different haptic vibration effects were added to Solar I3M. When a touch gesture is performed on a touch point an effect will trigger simultaneously. The Single Press feedback has a vibration duration of 150ms. Double Press activates two short vibrations of 150ms each, while Long Press vibrates for 500ms.}
\label{tab:table6}
\begin{tabular}{|c|c|c|c|}
\hline
\rowcolor[HTML]{EFEFEF} 
\textbf{Touch Gesture} & \textbf{Vibration Duration} \\ \hline
Activate/Single Press & 150ms \\ \hline
Double Press & 150ms (x2) \\ \hline
Long Press & 500ms \\ \hline
\end{tabular}
\end{table}

\subsubsection{Design Outcomes}
Across S2, S3 \& S4, our co-designers provided clear design feedback on how they wanted to engage in richer multimodal interaction. This included \textbf{using haptics to augment other modalities}, and facilitating non-visual experiences that holistically offer more meaning. \textbf{Natural interaction paradigms} were also desired, allowing users to engage in conversational dialogue using more human-like language, such as the use of demonstrative pronouns. 

\section{Evaluation}
All design outcomes observed throughout our co-design journey (Section \ref{sec:codesign}) were implemented into Solar I3M. We conducted a formal user study with blind participants in order to evaluate the design and usability of Solar I3M. Our evaluation consisted of a training and customisation period followed by both structured tasks and unstructured exploration in order to better understand the practical implications of our end-user-led design decisions.

\subsection{Participants}
Seven blind participants were recruited from our lab's participant contact pool. As a result of the low-incidence of blindness and associated difficulty in recruiting appropriate participants~\cite{Butler2021}, many studies in the blind accessibility space involve between 6-12 participants~\cite{Reinders2020,Shi2017b,Shi2020,Holloway2022,Nagassa2023}. Our work falls within this range. For user studies that follow co-design or participatory processes and involve blind users, it is not uncommon for evaluations to be conducted with even fewer participants~\cite{Ghodke2019,Giraud2020}. None of the participants recruited for the user study were involved in our co-design journey.

\begin{table}[!htbp]
\caption{Participant demographic information}
\Description{This table provides the demographic information of the seven participants that took part in Solar I3M's evaluation. All seven participants identified as blind. They used a range of accessible formats, including braille (6/7), audio (7/7), raised line (4/7) and models (4/7). Participants had varying levels of familiarity/confidence using tactile graphics, less so with tactile models and interactive models. Participants used conversational interfaces frequently, including Alexa (2/7), Bixby (1/7), Cortana (2/7), Google Assistant (6/7) and Siri (7/7). They accessed conversational interfaces most commonly on smartphones (7/7), followed by smart speakers (6/7), then tablets (2/7) and computer/television both on 1/7.}

\label{tab:table10}
\centering
\begin{tabular}{|l|c|c|c|c|c|c|c|} 
\hline
\cellcolor[HTML]{EFEFEF}\textbf{Participant} & \cellcolor[HTML]{EFEFEF}\textbf{P1} & \cellcolor[HTML]{EFEFEF}\textbf{P2} & \cellcolor[HTML]{EFEFEF}\textbf{P3} & \cellcolor[HTML]{EFEFEF}\textbf{P4} & \cellcolor[HTML]{EFEFEF}\textbf{P5} & \cellcolor[HTML]{EFEFEF}\textbf{P6} & \cellcolor[HTML]{EFEFEF}\textbf{P7} \\ 
\hline
\multicolumn{8}{|l|}{\cellcolor[HTML]{EFEFEF}\textbf{Level of Vision:}} \\ 
\hline
Blind & \checkmark & \checkmark & \checkmark & \checkmark & \checkmark & \checkmark & \checkmark \\  
\hline
Low-Vision & - & - & - & - & - & - & -\\ 
\hline
\multicolumn{8}{|l|}{\cellcolor[HTML]{EFEFEF}\textbf{Recent Accessible Formats Used:}} \\ 
\hline
Braille & \checkmark & \checkmark & - & \checkmark & \checkmark & \checkmark & \checkmark \\ 
\hline
Audio & \checkmark & \checkmark & \checkmark & \checkmark & \checkmark & \checkmark & \checkmark \\  
\hline
Raised Line & \checkmark & - & - & - & \checkmark & \checkmark & \checkmark \\  
\hline
Models & - & - & \checkmark & \checkmark & \checkmark & \checkmark & - \\  
\hline
\multicolumn{8}{|l|}{\cellcolor[HTML]{EFEFEF}\textbf{Familiarity (1: Not Familiar - 4: Very Familiar):}} \\ 
\hline
Tactile Graphics & 2 & 2 & 1 & 3 & 3 & 4 & 2 \\ 
\hline
Tactile Models & 1 & 1 & 1 & 1 & 2 & 2 & 2 \\ 
\hline
Interactive Models & 1 & 1 & 2 & 1 & 2 & 2 & 1 \\ 
\hline
\multicolumn{8}{|l|}{\cellcolor[HTML]{EFEFEF}\textbf{Recent Conversational Interfaces Used:}} \\ 
\hline
Alexa & - & - & - & \checkmark & \checkmark & - & - \\ 
\hline
Bixby & - & - & - & - & \checkmark & - & - \\ 
\hline
Cortana & - & - & - & - & \checkmark & \checkmark & - \\ 
\hline
Google Assistant & - & \checkmark & \checkmark & \checkmark & \checkmark & \checkmark & \checkmark \\ 
\hline
Siri & \checkmark & \checkmark & \checkmark & \checkmark & \checkmark & \checkmark & \checkmark \\ 
\hline
\multicolumn{8}{|l|}{\cellcolor[HTML]{EFEFEF}\textbf{Accessed on Devices:}} \\ 
\hline
Phone & \checkmark & \checkmark & \checkmark & \checkmark & \checkmark & \checkmark & \checkmark \\ 
\hline
Tablet & - & - & - & - & \checkmark & - & \checkmark \\ 
\hline
Speaker/Display & - & \checkmark & \checkmark & \checkmark & \checkmark & \checkmark & \checkmark \\ 
\hline
Computer & - & - & - & - & \checkmark & - & - \\ 
\hline
Television & - & - & - & - & \checkmark & - & - \\
\hline
\end{tabular}
\end{table}

All participants self-reported as blind (summarised in Table \ref{tab:table10}), with some reporting minimal low-levels of light perception. Age was spread evenly, from mid-20s to mid-70s. All participants regularly made use of text in the form of audio, while six knew how to read braille. Regular use of tactile models was not as common, however, six participants reported some or substantial past exposure to tactile graphics. In terms of technology use, all participants regularly used at least one conversational interface, with many reporting the use of several of these through smartphones and speakers. Common uses of conversational interfaces included checking the weather, dictating messages, and making phone calls.

\subsection{Procedure}
Each user study session lasted approximately 90 minutes, and began with a researcher providing a brief overview of the project, before guiding participants through the following components:
\begin{enumerate}
    \item \textbf{Training}. Participants were first asked to explore Solar I3M tactually with all system outputs turned off in order to build an initial level of comfort and confidence with the model. They were then taught to perform all of the touch gestures Solar I3M supported, as well as how to engage the conversational interface and supported voice commands. Participants were also guided through a series of sample interactions, and encouraged to perform their own interactions to build experience and comfort. This part of the session lasted 20-25 minutes.
    \item \textbf{Customisation}. Participants were then shown how they could customise Solar I3M's conversational outputs. Lasting 20-minutes, this included choosing a voice type \textit{[feminine, masculine, robotic]}, voice speed \textit{[100\%, 150\%, 200\%]} and response length \textit{[longer, shorter]}. For each customisable setting, participants were shown all available options, asked to choose their preferred setting, and informed that changes could be made at any time throughout the remainder of the user study.
    \item \textbf{Activity}. Once participants had built an understanding of how to use and customise Solar I3M, they completed one of two tasks: a set of five short structured activities, or an unstructured activity. Lasting a total of 10-15 minutes, the structured activities involved simple information gathering (e.g. discovering the largest planet, the temperature of Mars, etc). During the unstructured activity, participants were given up to 10 minutes to freely explore and interact with Solar I3M to discover and learn something previously unknown about the Solar System. The length of the study precluded participants from doing both, and separating them allowed us to compare participant behaviour.
\end{enumerate}

\subsection{Data Collection} We used two primary data collection methods, including a modified System Usability Scale (SUS) questionnaire, and a 30-minute semi-structured interview. Minor adjustments were made to the SUS questions, in consultation with our S3/S4 co-designer to ensure they would be meaningful in a blind context. All statements used in the SUS can be found in Appendix: Table \ref{tab:table11}. The semi-structured interview focused on participants' experiences with Solar I3M across key aspects of design and functionality, including touch, conversational, and multimodal interactions. Questions also explored whether Solar I3M's design respected participants' sense of agency, independence, and expectations around use. All participant sessions were video recorded for transcription.

\subsection{Analysis}
The goal of the user study was to assess the design and usability of Solar I3M from a qualitative perspective. We conducted a thematic analysis of: comments made by participants during the training, customisation and activity components; researcher observations of participant interactions throughout; and participant responses from the semi-structured interview. Based on our observations from the co-design sessions, we derived an initial set of codes and themes. Three members of the research team then independently coded one of the evaluation sessions, before meeting to extend the initial codes. All transcribed participant sessions were subsequently coded, with any conflicts between classifications reconciled in a subsequent meeting. Themes were then further refined and consolidated. 

\section{Results}
As alluded to earlier, our participants were split between those completing structured tasks (3 participants) and those engaging in unstructured exploration (4 participants). There were very minor differences in the types of interactions and strategies employed by the two groups. All participants started with a general idea of what they wanted to accomplish or discover, and chose a series of interactions that they perceived would be the most efficient means of accomplishing their goal. 

We first present results from the SUS questionnaire followed by participant outcomes with respect to touch, conversation, and multimodal interaction, as captured throughout all components of the evaluation. Each of these topics is presented from the point of view of our final themes, which include how \textbf{prior experience} with personal technology influenced participant expectations, how participants desired interactions that upheld their \textbf{independence and control} and respected their \textbf{confidence and comfort}, and the desire for \textbf{customisation and personalisation} of the model. We found the themes to be highly interlinked; for example, being able to customise the voice of the Solar I3M made participants comfortable, and this in turn positively affected their confidence in using the model.

\begin{figure}[h]
  \centering
  \includegraphics[width=1.00\columnwidth]{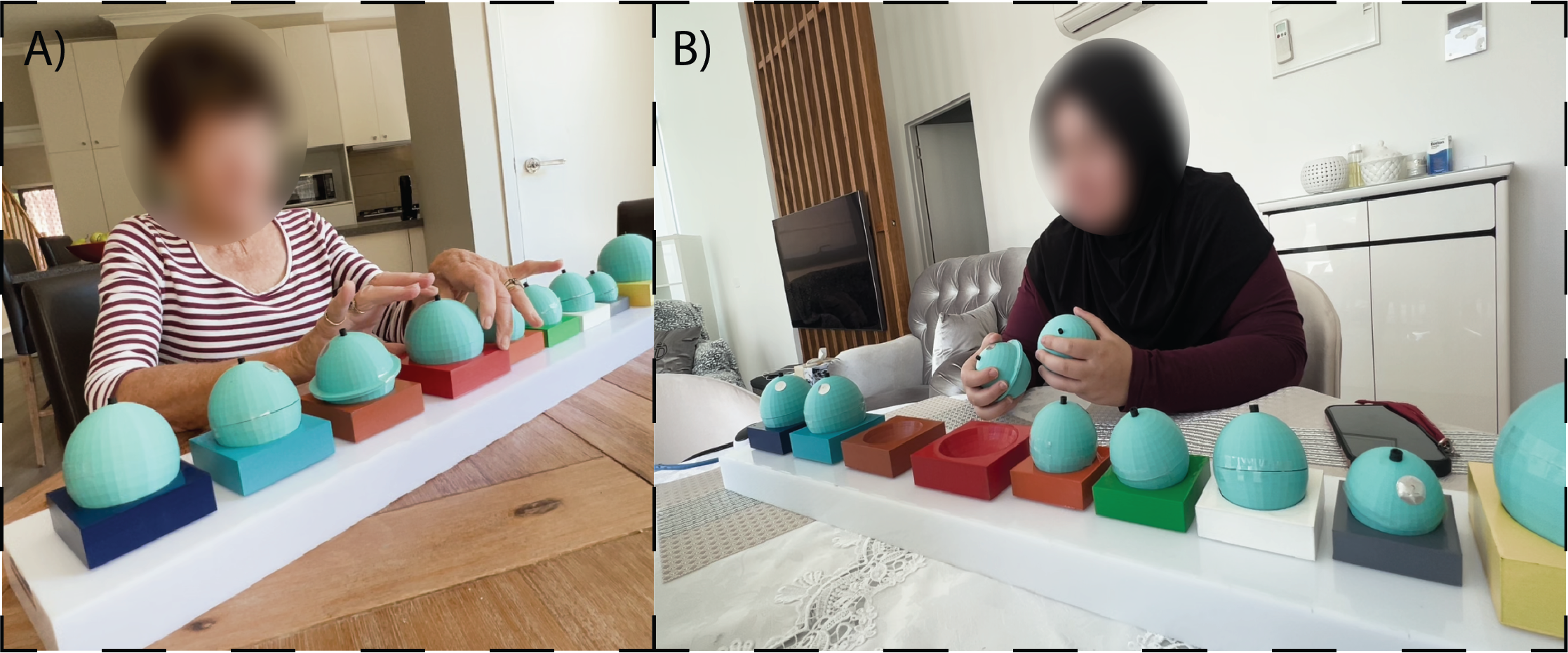}
  \caption{Participants interacting with Solar I3M, A) performing a Double Press touch gesture on Jupiter, and B) holding and tactually comparing Jupiter and Saturn while engaging in conversation with Solar I3M.}~\label{fig:eg-evlauation}  
  \Description{We showcase two photos of our participants performing different interactions with Solar I3M during the evaluation. The first photo, A), shows a participant bracing one hand against Jupiter and with the other hand performing a touch gesture on Jupiter's touch point. The second photo, B), shows a more complex interaction with a different participant holding the Saturn and Jupiter models, one in each hand. They are gathering detail tactually, and are about to ask Solar I3M a question comparing the planets.}
\end{figure}

\subsection{SUS Results}
Participants generally had overall positive experiences with Solar I3M, with an average SUS score of 84.29 (Table \ref{tab:table11}). All participants indicated that they would be interested in using I3Ms that represented other content (SUS1), and the consensus was that the use of similar I3Ms would improve participant's uunderstanding of the related concept (SUS5). Additionally, all felt comfortable when interacting with Solar I3M (SUS9). 
While participants generally found that it was easy to interact with Solar I3M (SUS3), some felt that they would be more comfortable interacting with I3Ms alongside a support person (SUS4). Upon further enquiry, P3 indicated that further experience and training might mitigate this need and could build up their comfort and familiarity, aligning with SUS10.

\begin{table*}[!h]
\caption{System Usability Scale responses}
\Description{This table lists the results from our System Usability Scale (SUS). With an overall SUS score of 84.29, The SUS statements and mean scores are provided below:
SUS1: I would like to use interactive models like this to access other content/concepts - Mean 4.86/5.00
SUS3: The interactive model was easy to use and interact with - Mean 4.00/5.00
SUS5: I think using interactive models like this would help me understand the relevant concept better - Mean 4.71/5.00
SUS7: I imagine that most people would learn to use interactive models like this very quickly - Mean 4.71/5.00
SUS9: I felt very confident and comfortable using the interactive model - Mean 4.71/5.00
SUS2: The interactive model was unnecessarily complex/difficult to use - Mean 1.71/5.00
SUS4: I think I would need the use of a support person to be able to use interactive models like this - Mean 2.29/5.00
SUS6: I think the information you can get with the interactive model is too limited - Mean 1.86/5.00
SUS8: I think the interactive model was cumbersome to use - Mean 1.71/5.00 
SUS10: I needed to learn a lot of things before I felt comfortable using the interactive model - Mean 1.71/5.00}
\label{tab:table11}
\centering
\begin{tabular}{|p{20mm}|p{20mm}|p{20mm}|p{20mm}|p{20mm}|c|}
\hline
\rowcolor[HTML]{EFEFEF} {\textbf{Str. Disagree}} & \multicolumn{1}{|>{\centering\arraybackslash}p{20mm}}{\textbf{Disagree}} & \multicolumn{1}{|>{\centering\arraybackslash}p{20mm}}{\textbf{Neutral}} & \multicolumn{1}{|>{\centering\arraybackslash}p{20mm}}{\textbf{Agree}} & \multicolumn{1}{|>{\centering\arraybackslash}p{20mm}|}{\textbf{Str. Agree}} & \multicolumn{1}{l|}{\textbf{Mean}} \\ 
\hline
\multicolumn{5}{|l|}{SUS1: I would like to use I3Ms like this to access other content/concepts} & \multirow{2}{*}{4.86} \\ \cline{1-5}
\multicolumn{1}{|c}{-} & \multicolumn{1}{|c}{-} & \multicolumn{1}{|c}{-} & \multicolumn{1}{|c}{1} & \multicolumn{1}{|c|}{6} & \\ \hline

\multicolumn{5}{|l|}{SUS3: The I3M was easy to use and interact with} & \multirow{2}{*}{4.00} \\ \cline{1-5}
\multicolumn{1}{|c}{-} & \multicolumn{1}{|c}{1} & \multicolumn{1}{|c}{1} & \multicolumn{1}{|c}{2} & \multicolumn{1}{|c|}{3} & \\ \hline

\multicolumn{5}{|l|}{SUS5: I think using I3Ms like this would help me understand the relevant concept better} & \multirow{2}{*}{4.71} \\ \cline{1-5}
\multicolumn{1}{|c}{-} & \multicolumn{1}{|c}{-} & \multicolumn{1}{|c}{-} & \multicolumn{1}{|c}{2} & \multicolumn{1}{|c|}{5} & \\ \hline

\multicolumn{5}{|l|}{SUS7: I imagine that most people would learn to use I3Ms like this very quickly} & \multirow{2}{*}{4.71} \\ \cline{1-5}
\multicolumn{1}{|c}{-} & \multicolumn{1}{|c}{-} & \multicolumn{1}{|c}{-} & \multicolumn{1}{|c}{2} & \multicolumn{1}{|c|}{5} & \\ \hline

\multicolumn{5}{|l|}{SUS9: I felt very confident and comfortable using the I3M} & \multirow{2}{*}{4.71} \\ \cline{1-5}
\multicolumn{1}{|c}{-} & \multicolumn{1}{|c}{-} & \multicolumn{1}{|c}{-} & \multicolumn{1}{|c}{2} & \multicolumn{1}{|c|}{5} & \\ \hline

\multicolumn{5}{|l|}{SUS2: The I3M was unnecessarily complex/difficult to use} & \multirow{2}{*}{1.71} \\ \cline{1-5}
\multicolumn{1}{|c}{3} & \multicolumn{1}{|c}{3} & \multicolumn{1}{|c}{1} & \multicolumn{1}{|c}{-} & \multicolumn{1}{|c|}{-} & \\ \hline

\multicolumn{5}{|l|}{SUS4: I think I would need the use of a support person to be able to use I3Ms like this} & \multirow{2}{*}{2.29} \\ \cline{1-5}
\multicolumn{1}{|c}{1} & \multicolumn{1}{|c}{4} & \multicolumn{1}{|c}{1} & \multicolumn{1}{|c}{1} & \multicolumn{1}{|c|}{-} & \\ \hline

\multicolumn{5}{|l|}{SUS6: I think the information you can get with the I3M is too limited} & \multirow{2}{*}{1.86} \\ \cline{1-5}
\multicolumn{1}{|c}{4} & \multicolumn{1}{|c}{1} & \multicolumn{1}{|c}{1} & \multicolumn{1}{|c}{1} & \multicolumn{1}{|c|}{-} & \\ \hline

\multicolumn{5}{|l|}{SUS8: I think the I3M was cumbersome to use} & \multirow{2}{*}{1.71} \\ \cline{1-5}
\multicolumn{1}{|c}{4} & \multicolumn{1}{|c}{2} & \multicolumn{1}{|c}{-} & \multicolumn{1}{|c}{1} & \multicolumn{1}{|c|}{-} & \\ \hline

\multicolumn{5}{|l|}{SUS10: I needed to learn a lot of things before I felt comfortable using the I3M} & \multirow{2}{*}{1.71} \\ \cline{1-5}
\multicolumn{1}{|c}{3} & \multicolumn{1}{|c}{3} & \multicolumn{1}{|c}{1} & \multicolumn{1}{|c}{-} & \multicolumn{1}{|c|}{-} & \\ \hline

\end{tabular}
\end{table*}

\subsection{Touch Interaction}
All participants were able to successfully engage with and perform touch interactions during their time with Solar I3M (Figure \ref{fig:eg-evlauation}a). Touch interaction generally transitioned from tactile exploration, assisting with immediate information gathering, to more comprehensive use of touch gestures to trigger audio label playback and invoke Solar I3M's conversational interface. As participants gained confidence in performing touch gestures, participants often picked up one or more of the planets to partake in more complex interactions.

\subsubsection{Prior Experience}
Participants drew heavily upon previous experiences with personal technology when performing touch interactions. This was seen across both \textit{tactile exploration} and when performing \textit{touch gestures}. All participants engaged in tactile exploration techniques when asked to explore Solar I3M. Participants more familiar with tactile graphics took part in more directed information gathering, while one participant with very limited experience -- \textit{P3} -- engaged in more limited exploration.

Solar I3M's touch gestures -- \textit{Single Press, Double Press} and \textit{Long Press} -- were largely understood by all participants. Participants drew strongly from experiences with personal technology, expecting the touch gestures to be designed largely analogous to use on smartphones and tablets. This led to some initial confusion as to how they were to be performed. \textit{P4} expected Single Press to act like the Single Tap gesture on their smartphone, describing \textit{``[the Single Press] was not intuitive for me... it is usually a quick tap... not a tap and hold''}. There was a clear understanding amongst participants about what each touch gesture was used for, with two participants directly connecting the use of the \textit{Long Press} to how they invoke the conversational interface on their smartphone (\textit{P4 \& P6}).

\subsubsection{Independence \& Control}
Participants agreed that the physical design of Solar I3M's touch points -- raised and protruding from the surface -- helped them to undertake independent tactile exploration and to easily identify what parts of the model they were able to perform touch interactions on, which directly aligned with our co-design finding. \textit{P4} suggested that touch points that were flush with the surface of a model could result in them \textit{``accidentally hitting something when [they weren't] ready''}, while \textit{P5} felt that the design of the touch points helped in orienting model components, \textit{``[the touch point] aids in placing the planet properly and making sure it is upright''}. One participant, \textit{P4}, advised that the design of the touch points could be further refined so that only the top would be sensitive to touch (pressable), finding that there were a few instances where they were able to accidentally trigger touch gestures by brushing the sides of the touch points.

The need for the Activate Press touch gesture was understood among participants. Three participants stated that the gesture made their touch interactions more deliberate (\textit{P3, P6 \& P7}), with \textit{P6} reporting that it gave them the control to \textit{``feel as though [they] wouldn't accidentally activate different planets ... unless it was explicitly what [they] wanted''}.

\subsubsection{Confidence \& Comfort}
Continued use influenced the confidence of participants and their willingness to perform touch interactions. While it was common for participants to experience slight confusion or hesitation when first introduced to the touch gestures, participants became more confident in their interactions over the course of the session. \textit{P3} spoke of how it took them time to \textit{``get used to how long or hard [they] had to press the touch points''}, while \textit{P5} suggested that a practice mode similar to the VoiceOver screen reader on their smartphone would increase their confidence if they were to encounter similar I3Ms in day-to-day life. One participant, after experiencing difficulty reliably performing the Single Press gesture early on during their session, was even confident enough to state towards the end of the session that \textit{``with use and with regularity ... the presses could take over as [my] preferred way [to interact]''} (\textit{P7}).

\subsubsection{Customisation \& Personalisation}
During the user study sessions a number of participants emphasised a desire to customise how touch interactions could be performed. Despite being able to reliably perform the touch gestures, \textit{P6} suggested that the sensitivity of the touch gestures should be customisable and that this would be particularly useful for users who might have different levels of dexterity. Recalling their smartphone usage, \textit{``on Android you can change the tap sensitivity ... with certain mobility issues, [touch gestures] would be trickier [to perform]''} (\textit{P6}). \textit{P4} desired control over the timing windows of touch gestures in order to make their interactions quicker and more efficient -- e.g. being able to make \textit{Single Press} faster and more analogous to the use of the Single Tap on their smartphone. 

\subsection{Conversational Interactions}
Participants were largely comfortable engaging with the conversational interface from the moment it was introduced. Throughout the course of the user study, participants began gravitating more towards this form of interaction, which was seen as faster and more efficient. 

\subsubsection{Prior Experience} 
The expectations that participants had when undertaking conversational interactions were largely influenced by their experiences with the conversational interfaces found on their personal technology. This extended to how participants \textit{expected to be able to engage} with Solar I3M's conversational interface. 

The use of a \textit{wake word} --  \textit{``Hey Model''} -- was well understood amongst participants. This specific phrase was seen as being analogous to phrases used on their smart devices, \textit{``well this is sort of going back to Apple and Siri... [Hey Siri]''} (\textit{P7}). Using the \textit{Long Press} touch gesture as a method of physically waking the conversational interface was also well understood. \textit{P4} directly spoke of how a similar touch gesture was used on their smartphone, \textit{``[this] is how you manually wake up voice assistants, there is continuity there''}. 

\subsubsection{Independence \& Control} 
The conversational interface was seen as being the most efficient method of interaction among participants. Participants felt that engaging in a conversational interaction would be more straightforward, giving them the greatest level of control. The \textit{``Hey Model''} wake word was also considered to be the fastest way of invoking the conversational interface, with \textit{P7} mentioning how it was quicker and more direct. \textit{P4} opted to use the wake word because they felt it gave them more control, finding the \textit{Long Press} gesture to be \textit{``arduous ... [it] was too long''}.

Despite not actively using it, all participants agreed that it was imperative to have a voice command that could repeat auditory responses -- \textit{``can you repeat that?''}. This directly validated our co-design finding. \textit{P4} even suggested this command before it was covered in the user study, \textit{``there is nothing as frustrating as if I can't have it repeat what is said... then I have to cycle through things... or just ignore it''}.

Participants felt that being able to seek responses from Google Knowledge Graph was particularly empowering, allowing them to undertake more extensive interactions. One participant, \textit{P3}, spoke of how when using a conversational interface on their personal technology, they would often experience frustration when it responded with the message -- \textit{``I don't have any information on that''}. 
The majority of participants felt in control when Solar I3M entered its search flow -- \textit{``I don't know the answer to [question], but would you like me to search for it?''}. \textit{P6} described how it made their interactions with Solar I3M \textit{``[feel] more like a conversation rather than just using my voice as a keyboard''}. 

A major area of friction during conversational interactions included the support of limited-phrase variations in voice commands. This often resulted in participants uttering voice commands that went undetected by Solar I3M, causing confusion. This occurred when Solar I3M would prompt participants if it should enter its search flow. Expecting the response -- \textit{``Yes, search for it''} -- some participants would answer -- \textit{``Yes''} -- which would go undetected. \textit{P2} discussed how voice commands should support more phrase-agnostic expressions, saying that \textit{``there is a lack of intuitiveness when you [say] `yes go ahead, look it up', I want the process to be simpler... why not just yes''}.

\subsubsection{Confidence \& Comfort} 
Participants were comfortable asking questions using the conversational interface. This was no surprise to the researchers, as all seven participants indicated regular use of conversational interfaces on their personal technology (Table \ref{tab:table10}). \textit{P5} openly indicated their comfort engaging in conversational interactions based on their prior experience, \textit{``honestly because I am accustomed to using smart devices like Alexa, it comes naturally to me to ask questions''}. \textit{P4} outlined that they would not be comfortable engaging in conversational interaction in public spaces, in part to not inconvenience others and due to privacy, describing \textit{``in this [environment] where it is private, I will speak to it, but in more public environments, I prefer physical touch interactions''}. \textit{P4} suggested that a headphone option might help somewhat mitigate this, but still felt uncomfortable engaging using speech and conversation open in public.

\textit{P7} described how when the I3M offered to perform a search, having it admit that it couldn't answer them directly -- \textit{``I don't know the answer to [query/question]...''} -- increased their confidence and trust, describing it as \textit{``being genuine... it enhanced my confidence, because it was aware of what it knew and what it [didn't]''}. \textit{P6} elaborated further, \textit{``It helps build trust, rather than the teacher just making something up, it admits that it is fallible... it makes it seem more genuine''}. All seven participants emphasised that providing references was critical to their comfort, allowing them to come to their own interpretations as to how trustworthy returned information was. This directly aligned with one of our co-design findings.

\subsubsection{Customisation \& Personalisation}
The ability to modify conversational output was seen as critical amongst participants, allowing them to reduce friction and consume information in a way that suited them. 
All participants were able to successfully use voice commands to choose their preferred voice type, speed, and response verbosity. 
Individual preferences were crucial, \textit{``what suits one person may not suit another... but what I have chosen was good for me''} (\textit{P3}). Participants also described how their preferences could change based on multiple variables, including:

\begin{itemize}
\item \textbf{Their own \textit{ability}:} Participants said that differing levels of hearing and comprehension could greatly influence how they customised conversational output. \textit{P4 \& P2} indicated that due to mild hearing impairment, they found human voices easier to understand compared to highly synthesised voices, while \textit{P7} outlined that a memory impairment meant that they wouldn't confidently be able to follow at speeds greater than 100\%.

\item \textbf{Their own \textit{personal preference}:} Participants indicared that they chose voices based on pleasantness, with more human-like WaveNet voices having more satisfying inflection and intonation. \textit{P6} preferred WaveNet voices, but suggested that they might choose the eSpeak voice on their screen reader of its familiarity.

\item \textbf{The \textit{environment} where the I3M is being used:} \textit{P6} said that in loud environments -- e.g. a busy train -- they might decrease the speed of conversational output so that they can better interpret speech, and also suggested that in a quiet environment, they might increase the speed and use a more synthetic-sounding robotic voice. 

\item \textbf{The \textit{content} the I3M represents:} \textit{P1, P2, P3 \& P5} discussed how if the I3M represented content that they were particularly interested in learning about, they might prefer slower responses that were more easily consumable, with \textit{P5} indicating that \textit{``if you want to enjoy something and think about it deeply, then the speed needs to be at the pace of your thought process''}. \textit{P6} said that for more technical and difficult-to-digest content -- e.g. university materials --- they would prefer the more verbose output.

\item \textbf{The \textit{purpose} the I3M is being used for:} \textit{P1, P4, P5 \& P7} indicated that if they were using an I3M to revise information they had already learnt, they might increase the speed of the conversational output, or reduce the response verbosity. \textit{P4} spoke of how with critical tasks they might make output slower and more verbose in order to focus, \textit{``if is it important for my life, like a new environment I want to physically go to, I would change it''}.
\end{itemize}

Participants were eager to make additional voice configuration suggestions.
In particular, \textit{P6} requested a voice type that was gender-neutral, in part because they felt their preference for the feminine voice might have been based on gender stereotypes, describing \textit{``there might be some bias there... like the female secretary type thing... I don't know''}.

\subsection{Multimodal Interactions}
Throughout the user study, participants seamlessly performed interactions involving combinations of auditory and haptic input/output. It was not uncommon for participants to pick planets up, perform touch gestures on them whilst being held, and then engage in conversational interactions. Participants found the use of haptic vibratory feedback and demonstrative pronouns appealing, creating more natural experiences that helped to bind their interactions together. Amongst some participants, there was a desire for interactions and experiences with I3Ms that would extend further into the space of more physically embodied models.

\subsubsection{Prior Experience}
Participants weren't explicit about whether any of their previous experiences with personal technology influenced their interaction strategy or willingness to combine and switch between modalities during interactions. It is, however, our belief that participants may have unknowingly been influenced by their use of personal technology, specifically smartphone use. When asked about their strategy, \textit{P6} described their use of different modalities as \textit{``...just [doing] what came to mind''}. We feel that the touch reading experience participants had also likely played a role in their willingness to engage and combine tactile information gathering and more directed touch -- e.g. picking planets up -- during their interactions.

\subsubsection{Independence \& Control}
Participants expressed that elements of Solar I3M's multimodal design helped them to stay in control and independent during interactions. This was seen by how \textit{adaptable} Solar I3M was, and how haptic vibratory feedback was used to \textit{augment interactions} by providing added feedback and meaning. Participants provided direct suggestions on how vibratory feedback could further augment interactions to enhance their sense of control. 

Participants found it empowering being able to choose their preferred interaction modalities, and how they could extract and access the same information using their choice of touch gesture input or conversational interaction. Participants began describing different situations where they might shift between the two, including based on the environment or the task. \textit{P7} spoke of how they might choose to use touch gesture input to trigger audio label playback because they perceived that it was quicker than engaging in conversation, while \textit{P5} said that if they were using an I3M to introduce themselves to a subject, they wanted to \textit{``immerse [themselves] the way they would with a tactile image''}. \textit{P4} indicated that the environment could be a determining factor, discussing how \textit{``in this [environment] where it is private, I will speak to it, but in more public environments, I prefer physical touch interactions''}. Participants emphasised that it was vital that they weren't made to feel that they were missing out on information based on the interaction modality they chose to use -- e.g. if touch gesture audio label playback gave shorter responses compared to conversational output.

There was agreement amongst all participants that the use of haptic vibratory feedback to augment touch and conversational interactions helped to increase a sense of control during interactions. Aligning directly with our co-design finding, participants found that haptic vibratory feedback helped to confirm their touch gesture inputs, \textit{P3} described \textit{``you knew that you'd done the right thing ... Single Press, Double Press, Long Press ... [haptics] confirmed it... ''}. \textit{P5} was more direct about how it improved their sense of control, speaking \textit{``if I didn't have the feedback, I would keep [repeatedly] pressing''}.

\subsubsection{Confidence \& Comfort}
During interactions, participants were seen comfortably changing between and combining different modalities, perceiving them as creating a \textbf{more natural experience}. 

Participants enjoyed referring to planets using demonstrative pronouns during interactions. One participant, \textit{P5}, was particularly fascinated by the concept, and repeatedly performed comparisons while holding a planet in each hand -- asking questions including \textit{``how large are \textbf{these}?''} and \textit{``which of \textbf{these} is closer to the Sun?''} (Figure \ref{fig:eg-evlauation}b). When asked, \textit{P5} suggested that using pronouns made them more willing to pick planets up as part of their interactions, and that it made them more confident to explore and undertake more complex interactions. This aligned with the experiences of our co-designers.

Following on, \textit{P5} was so comfortable during their interactions with Solar I3M that they began suggesting feedback that would allow them to partake in richer natural interactions with a more embodied Solar I3M. Asking if auditory output could be emitted from a speaker embedded directly within the planet they were holding, \textit{P5} described that \textit{``it would change the experience, right now they feel like spheres with no life other than when they vibrate, information just comes out of a speaker''}. Further elaborating, they suggested that a personality or character could be imbued within Solar I3M, detailing \textit{``... or a guide... Einstein? ... [or] I am the Sun, be careful when you pick me up, I might be too hot!''}.

\subsubsection{Customisation \& Personalisation}
While all participants found the use and combination of haptic vibratory feedback during touch and conversational interactions valuable, there was a desire amongst some participants to have more control over how it was used. One participant spoke of how due to sensory considerations, some users may want to customise any haptic vibratory feedback, specifically changing the intensity and length of the vibrations or even turning them off altogether (\textit{P4}). Despite classifying themself as a \textit{``very pro-haptics person''}, they further described how during one interaction, they wanted to interact focusing solely on Solar I3M's auditory responses, uninterrupted by any haptic vibratory feedback. \textit{P4}'s response perhaps suggests that when it comes to modalities used to augment I3M outputs, some blind users may need better support to choose how multimodal or unimodal the output will be.

\section{Discussion}
The role of personal technology played a significant factor during Solar I3M's design and evaluation. Specifically, our blind co-designers and participants desired interactions that drew from built-up knowledge from their \textit{prior experience} with personal technology -- e.g. smartphones. This applied to the touch gestures users could perform on Solar I3M and how the conversational interface operated, helping to increase the \textit{confidence and comfort} of our users. The use of touch gestures is well understood amongst blind users due to the proliferation of touch input devices, and prior work has identified that blind users desire I3Ms that support consistent touch-based experiences~\cite{Shi2017a,Shi2017b}. However, the degree to which prior experience influences conversational interfaces is less clear. ~\cite{Reinders2020} observed participants using smart speaker-analogous wake words with a WoZ I3M, but found that built-up experience with conversational interfaces seemingly didn't influence the likelihood of users engaging in interactions with conversational dialogue. Our observations deviated from this, with participants finding the conversational interface the most efficient way of interacting with Solar I3M, drawing a direct connection to their use of smart devices. One possible reason for this might be that since this work in 2020, blind users may have become more reliant on and comfortable with conversational interfaces. It is clear that going forward I3Ms should be designed to be consistent with interaction experiences available on personal technology, allowing blind users to transfer knowledge.

The use of Solar I3M also contrasted with our previous work involving a WoZ I3M of the Solar System, where participants employed an interaction hierarchy: from tactile exploration to touch gesture audio labels and natural language questioning. Our findings suggest that the choice of interaction strategy may be more nuanced, with other variables impacting how blind users choose to interact with an I3M -- including \textit{task, environment, and familiarity with tactile graphics and conversational interface} -- e.g. one participant was adamant that they wouldn't engage in any conversational interactions in public.

Our co-designers and participants provided design feedback on how Solar I3M could better support their \textit{independence and control}. The need for independent experiences has been identified in prior work, allowing blind users to use I3Ms without the need of support workers~\cite{Shi2019}, and to come to their own interpretations~\cite{Reinders2020}. ~\cite{Holloway2018,Reinders2020} both observed blind users becoming frustrated when interrupted by I3M audio label playback during tactile exploration. Our findings aligned with these works, providing direct feedback on the physical design of touch points and activation gestures. But as Solar I3M is also the first I3M that integrates a fully-functional conversational interface inside the model itself, we also observed that the desire for independence extends beyond touch to conversational dialogue, and gathered direct design feedback that supports the independence of blind  users when using Solar I3M to ask questions and perform searches.

The possibilities of multimodal interactions informed Solar I3M's design. Whereas most I3Ms combine printed models with auditory output in a limited fashion~\cite{Giraud2017,Gotzelmann2017,Holloway2018,Ghodke2019}, Solar I3M involvedtightly integrated conversational dialogue, audio, haptic vibration, and touch during interactions. Our co-designers directly suggested ways in which haptic vibratory feedback could be used to augment auditory responses with additional context/meaning, and confirm the successful activation of their touch gestures. We also explored the use of demonstrative pronouns, markedly extending previous work in accessibility research which used conversational interfaces in service of voice-driven menu navigation and extraction of audio labelling~\cite{Bartolome2019,Quero2018}. Demonstrative pronouns were integral to facilitating blind participants' engagement in more natural dialogue during interactions, leading one participant to desire I3Ms with more physically embodied characters. 

The ability to \textit{customise and personalise} Solar I3M was strongly desired amongst our participants during their interactions. This extended across all aspects of Solar I3M's design: from touch gestures, to the conversational interface, and to associated auditory output and haptic vibratory feedback. Our participants indicated that being afforded this level of control would allow them to determine their own pace during interactions, and that preferences could change based on multiple variables, including ability and environment. Prior work has already identified that customisation of conversational interfaces is needed to help remove restrictions on blind users' interactions~\cite{Abdolrahmani2018,Branham2019,Choi2020}. Our findings align with this work, highlighting the importance of customisation, but extending to a new context -- when conversational interfaces are integrated inside tangible I3Ms. We also observed that our users expected to be able to customise other interactions, including how touch gestures were performed, and the timing, sensitivity and intensity of haptic vibratory feedback. To our knowledge, this level of I3M customisation has not been previously considered, and highlights a need to support blind users in customising interactions to cater to very personal preferences and abilities.

\subsection{I3M Design Recommendations} 
We developed a set of design recommendations based on our observations and the results from Solar I3M's design journey and evaluation. These recommendations coalesce across our identified themes, and focus on the design of I3Ms that support blind users' desire for -- use of \textbf{prior experience}, interactions that uphold their \textbf{independence and control}, sense of \textbf{confidence and comfort}, and allow \textbf{customisation and personalisation}. The following design recommendations are being put forward to the accessibility research community to help guide the implementation of future I3Ms:

\begin{itemize}
    \item \textbf{\textit{Interruption-free tactile exploration:}} I3Ms should be designed to respect the \textit{independence and control} of blind users, especially when any tactile exploration takes place to discover/interpret information. Touch points should be physically designed to be easily identifiable, and may protrude or recess~\cite{Holloway2018} the surface of the model based on the degree of salient details on the model. Additional features, including an Activate Press touch gesture, on/off buttons~\cite{Shi2017b}, or voice commands, should be used to prevent unintended touch or conversational interactions from occurring, reducing friction for blind users during initial exploration or subsequent interactions.

    \item \textbf{\textit{Leverage prior interaction experience with personal technology:}} Blind users expect to be able to interact with I3Ms in ways similar to their smartphones and smart speakers, aligning with ~\cite{Reinders2020}. I3Ms should be designed to take advantage of this \textit{prior experience}, supporting interactions that are similar to those used with personal technology. Users should be able to perform gestures -- including Single, Double and Long Taps or Presses -- during touch interactions to extract basic information. Users should also be able to use wake words, touch gestures, and easy-to-remember voice commands -- similar to Siri and Google Assistant -- when engaging in conversational interactions to extract richer information that fills gaps in knowledge. 

    \item \textbf{\textit{Support customisation and personalisation:}} I3M interactions should be designed to be customised, allowing blind users to determine their own pace and preference when performing interactions. Based on personal preference, task or environment, \textit{customisation and personalisation} should extend to how users both perform inputs and receive output during their interactions, influencing their \textit{confidence and comfort}. Touch gestures should have adjustable timing and sensitivity, while haptic vibratory effects should have adjustable intensity and timing, or be able to be turned off to support any users with sensory considerations. Conversational output should have customisable high-quality voice types, voice speeds, and response lengths~\cite{Abdolrahmani2018,Branham2019}, allowing blind users to determine the pace and form in which auditory responses are consumed.

    \item \textbf{\textit{Support more natural dialogue:}} I3Ms should allow blind users to leverage conversational interfaces to fill gaps in their knowledge, empowering their desire for \textit{independence and control} during interactions. Users should be able to use demonstrative pronouns~\cite{Bolt1980}, including \textit{``these, this, it, that''}, to refer to any I3M model components being touched or held. 
    Demonstrative pronouns help support more intuitive dialogue, which is particularly important for blind users, allowing them to engage conversationally with objects that are yet to be verbally identified, versus a sighted user who might identify an object using visual information (e.g. colour). Additionally, users should be able to authorise I3Ms to perform searches and be able to respond to follow-up queries using straightforward language -- e.g. command-agnostic responses \textit{``yes'', ``okay'', ``sure''} over command-specific phrases like \textit{``yes, search for it''} -- increasing their \textit{confidence} and \textit{comfort} when engaging conversationally.

    \item \textbf{\textit{Tightly coupled haptic feedback:}} Haptic vibratory feedback should augment speech and touch to create more intuitive and natural interactions. It can be used to confirm the successful activation of touch gestures, increasing the \textit{independence} and \textit{control} of blind users as they validate that their interactions are being fulfilled. Haptic vibratory feedback localised within model components allows blind users to better relate conversational responses to specific components that are being held, compared to sighted users who are able to use visual information to help achieve this. In general, tightly coupled haptic feedback supports richer interactions -- e.g. comparing multiple components held simultaneously, and could be used with an I3M of a map in order to more effectively direct a user to a destination using combinations of audio and haptic feedback. 
\end{itemize}

\section{Limitations \& Future Work}
Solar I3M's design and evaluation resulted in the creation of our design recommendations. However, a limitation is that our recommendations were distilled from the design of a single I3M. A major focus of our future work will be to examine and expand these using different models -- e.g. an I3M map used for orientation and mobility training. Additionally, despite seeing a large degree of consistency across our participant group, we would like to run evaluations with a larger participant base. We would also like to run a similar study with low-vision participants, as we suspect the tactile and conversational nature of Solar I3M will be useful in low-vision contexts. These will add value and strengthen the generalisability of our recommendations.

We also aim to explore the design of more physically embodied I3Ms. Some of our participants desired interactions that were more personal and \textit{alive}. This included wanting to address conversational commands directly to physical components being held (e.g. \textit{``Mars how big are you?''}), having auditory responses delivered from speakers inside individual components rather than an enclosure/stand, and deeper use of haptic vibratory feedback. This would be particularly interesting in that the embodiment of interactive interfaces and agents traditionally relies on the perception of many visual attributes~\cite{Shamekhi2018}. I3Ms could have modes further along the human-to-human conversational model, introducing additional conversational protocols~\cite{cassell2000more} and a more human-like social presence~\cite{Luria2019}.

We are also interested in investigating the role of trust between blind users and I3Ms. Trust has proven to be a significant factor in other under-served populations when deploying new systems (e.g. the elderly in healthcare~\cite{Spillane2019,Carros2020}). Developing a trust relationship is particularly important for people who are blind, as the usefulness of an accessible graphic, aid, or tool depends entirely on whether the user is willing to accept and engage with the information it provides. With Solar I3M, the stakes were relatively low; if incorrect information was provided, users could merely ignore it or cease using Solar I3M. However, an opportunity exists to explore the concept of trust further, particularly with I3Ms in contexts where trust and confidence are more significant -- e.g. with orientation and mobility training, in which erroneous information could have safety implications.

\begin{acks}
This research was supported by an Australian Government Research Training Program (RTP) Scholarship. We wish to thank our co-designers and participants for their time and expertise.  
\end{acks}



\bibliographystyle{ACM-Reference-Format}
\bibliography{references}


\end{document}